# Performance of volume phase gratings manufactured using ultrafast laser inscription

David Lee[*a], Robert R. Thomson[b], and Colin R. Cunningham[a]
[a] Science and Technology Facilities Council, UK Astronomy Technology Centre,
Royal Observatory, Blackford Hill, Edinburgh, EH9 3HJ, U.K.
[b] Scottish Universities Physics Alliance (SUPA), Institute of Photonics and Quantum Sciences,
School of Engineering and Physical Sciences, David Brewster Building, Heriot Watt University,
Edinburgh, EH14 4AS, U.K.

**ABSTRACT**

Ultrafast laser inscription (ULI) is a rapidly maturing technique which uses focused ultrashort laser pulses to locally modify the refractive index of dielectric materials in three-dimensions (3D). Recently, ULI has been applied to the fabrication of astrophotonic devices such as integrated beam combiners, 3D integrated waveguide fan-outs and multimode-to-single mode convertors (photonic lanterns). Here, we outline our work on applying ULI to the fabrication of volume phase gratings (VPGs) in fused silica and gallium lanthanum sulphide (GLS) glasses. The VPGs we fabricated had a spatial frequency of 333 lines/mm. The optimum fused silica grating was found to exhibit a first order diffraction efficiency of 40 % at 633 nm, but exhibited approximately 40 % integrated scattered light. The optimum GLS grating was found to exhibit a first order diffraction efficiency of 71 % at 633 nm and less than 5 % integrated scattered light. Importantly for future astronomy applications, both gratings survived cooling to 20 K. This paper summarises the grating design and ULI manufacturing process, and provides details of the diffraction efficiency performance and blaze curves for the VPGs. In contrast to conventional fabrication technologies, ULI can be used to fabricate VPGs in almost any dielectric material, including mid-IR transmitting materials such as the GLS glass used here. Furthermore, ULI potentially provides the freedom to produce complex groove patterns or blazed gratings. For these reasons, we believe that ULI opens the way towards the development of novel VPGs for future astronomy related applications.

## 1. INTRODUCTION

Ultrafast Laser Inscription (ULI) is a process which uses ultrashort pulses of light to directly inscribe complex three-dimensional modifications into transparent dielectric materials. The process uses a focussed high peak power, ultrashort pulsed laser beam to modify the refractive index of glass in the region where the laser beam is brought to a focus (Figure 1). ULI has been successfully used to manufacture a number of photonics devices, such as multimode-to-single-mode integrated photonic lanterns [1], integrated beam combiners for long-baseline interferometry [2], and 3D fan-out devices [3] which are amongst the building blocks of astrophotonic instrumentation [4]. However, the ULI process also provides the capability to manufacture more traditional types of optics such as diffraction gratings.

The performance advantages of volume phase holographic gratings, which contain a holographic grating structure recorded in a gelatin layer, are well known [5] but the use of gelatin restricts the wavelength of operation to below approximately 2.2 μm. The ULI process allows gratings to be manufactured in many different types of material [6][7] and this potentially enables the production of volume phase gratings that operate over wavelength ranges from the visible to the mid-infrared. The use of ULI to manufacture grating structures therefore offers, *for the first time*, the possibility to produce a high efficiency volume phase grating that operates at mid-infrared wavelengths. The ULI process allows the grating to be manufactured directly in a mid-infrared transmitting material by direct-writing, avoiding the need for holograms. The grating produced will be also very robust as there are no separate layers as there are with holographic gratings, just one piece of glass material.

---

[*] Contact e-mail: david.lee@stfc.ac.uk

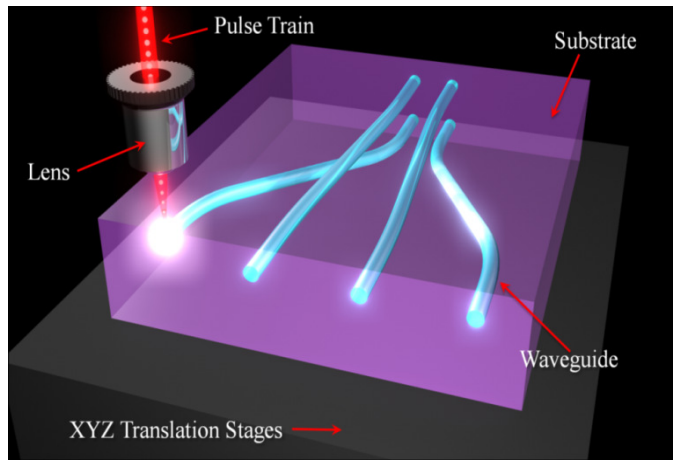

Figure 1. Schematic of the ULI manufacturing process.

The complete three-dimensional freedom provided by ULI processing also has the potential to produce complex grating structures that cannot easily be obtained by holographic techniques. For example volume phase gratings with variable line spacing, blazed profiles, or curved lines.

The Science and Technology Facilities Council UK Astronomy Technology Centre and Heriot-Watt University have recently embarked on a collaborative project to investigate the potential of using ULI for the manufacture of volume phase grating structures for applications in astronomy and remote sensing. This paper reports on preliminary results obtained from prototype volume phase gratings manufactured in Fused Silica and Gallium Lanthanum Sulphide substrates. The next section provides a brief review of ULI components and previous attempts at grating manufacture. The ULI manufacturing process is then described in more detail. The results of a series of optical performance tests are then described followed by a summary, conclusions, and an overview of potential future work.

## 2. REVIEW

The first demonstrations of using ultrashort pulses of light to inscribe 3D refractive index structures inside transparent materials were made in 1996 in two seminal papers [8][9]. Since then, the technique has rapidly developed, and is now finding applications in areas ranging including microsystems [10], biophotonics [11], waveguide lasers/amplifiers [12][13] and now astrophotonics for the development of multimode-to-single-mode integrated photonic lanterns [1], integrated beam combiners for long-baseline interferometry [2], 3D fan-out devices [3] and integrated optical waveguide pupil remappers [14].

It is believed the first diffraction grating structures manufactured using ULI were in fused silica [15][16]. One grating [16] had a peak efficiency of 20 % and the efficiency was found to be dependent on the polarisation of the writing laser beam and the measurement beam. Further work on manufacturing gratings in fused silica [17] describes a series of 0.3 mm by 0.3 mm prototype gratings with periods of 3 μm, 4 μm, and 5 μm and depths of 33 μm, 150 μm and 200 μm. The maximum observed diffraction efficiency was 75 % for TE polarised light and 59 % for TM polarisation. From these measurements they calculate, via Kogelnik's coupled wave theory, a refractive index modulation of 0.002 – 0.01 but do not measure the modulation directly. A further ULI grating inscribed in Schott filter glass type OG530 [18] produced a large refractive index modulation of 0.018. This value was measured directly using a refracted near field profilometer. The OG530 gratings had a peak efficiency of 37 %. Although not mentioned in the paper the use of filter glass to manufacture gratings provides the potential advantage of built in blocking of unwanted wavelengths. Volume phase gratings have also been manufactured in Foturan photosensitive glass [19] with peak efficiency of 56 % but the refractive index modulation is not mentioned.

This previous work has demonstrated that gratings can be manufactured using ULI, in a variety of materials, but many of the results are rather vague and no one has yet presented a rigorous study of the efficiency performance of ULI gratings. The grating parameters of key interest are the refractive index modulation achieved with ULI, efficiency and scattered light performance. These key details, measured for this work, are presented in Table 3.

## 3. GRATING MANUFACTURE

Our gratings were fabricated via ULI, using an ultrafast Yb-doped fiber laser system (IMRA FCPA μ−Jewel D400). The system supplied 340 fs pulses of 1047 nm radiation at a pulse repetition frequency of 500 kHz. The substrate material was mounted on x-y-z air−bearing translation stages (Aerotech ABL1000) which facilitated the smooth and precise translation of the sample through the laser focus. For the purposes of our work, we fabricated the volume gratings by repetitively scanning the substrate back and forth through the laser focus. Each full grating was composed of a number of layers of index modification - each stacked one on top of another. The laser focus was created by focusing the laser using an aspheric lens with a numerical aperture of 0.4, resulting in a spot size (diameter) of approximately 1.5 μm. In this configuration the ULI process is most well suited to manufacturing volume phase gratings with relatively low line densities < 500 lines per mm. However, we note that other techniques, such as the use of phase masks to define the irradiation pattern, do exist and can facilitate much higher groove densities [6]. In comparison the interferometric technique used to manufacture volume phase holographic gratings is best suited to high line densities [5] and cannot easily achieve the low line density suited to the ULI technique. A grid of 16 prototype gratings was produced on each substrate (Figure 3). The fabrication parameters for each grating were identical, except that we varied to the laser power in order to explore the effect of pulse energy on the grating performance. Table 1 summarises the design parameters of the prototype gratings and Table 2 summarises the parameters used to inscribe the gratings.

Table 1. Summary of the parameters of the prototype gratings.

| Parameter | Specification |
| --- | --- |
| Substrate materials | Fused Silica and Gallium Lanthanum Sulphide (GLS) |
| Substrate dimensions | 25 mm diameter, 1 mm thick |
| Number of prototype gratings | 16 arranged in a 4 by 4 grid |
| Grating size | 3 mm by 3 mm |
| Grating spacing | 1 mm between gratings on substrate |
| Line density | 333 lines per mm |

The materials chosen for the ULI volume phase grating prototypes were fused silica and gallium lanthanum sulphide. Fused silica was chosen because the ULI processing steps for this material are well understood. GLS was chosen for two reasons: its ability to produce a relatively high refractive index modulation and its good optical transmittance from approximately 0.5 – 10 μm, with its potential application to infrared astronomical spectroscopy.

Table 2. Summary of the ULI manufacturing parameters.

| Parameter | Specification |
| --- | --- |
| Laser | IMRA FCPA μ−Jewel D400 |
| Pulse repetition rate | 500 kHz |
| Pulse duration | ≈340 fs |
| Polarisation | Circular |
| Pulse energies | 640 nJ → 80 nJ (Fused silica) / 180 nJ → 10 nJ (GLS) |
| Scan speed | 10 mm per second |
| Objective NA | 0.4 |
| Number of scan layers | 17 |
| Scan separation | ≈ 3 μm (Fused silica) / ≈ 2.4 μm (GLS) |

The ULI process produces a region of modified refractive index, the depth of which is approximately related to the confocal parameter of the focused beam ($b$), where $b = (2\pi \omega^2/\lambda)$. We therefore estimate that the depth of the modified region will be ≈ 6 μm and 10 μm for the case of silica ($n ≈ 1.5$) and the GLS ($n ≈ 2.4$) respectively. For a volume phase grating to operate efficiently at a wavelength 633 nm, assuming a pitch of 3 μm and a refractive index modulation of

0.005, requires a grating depth in excess of 30 μm. The grating depth was therefore built up by creating multiple planes of gratings, each stacked on top of one another [18].

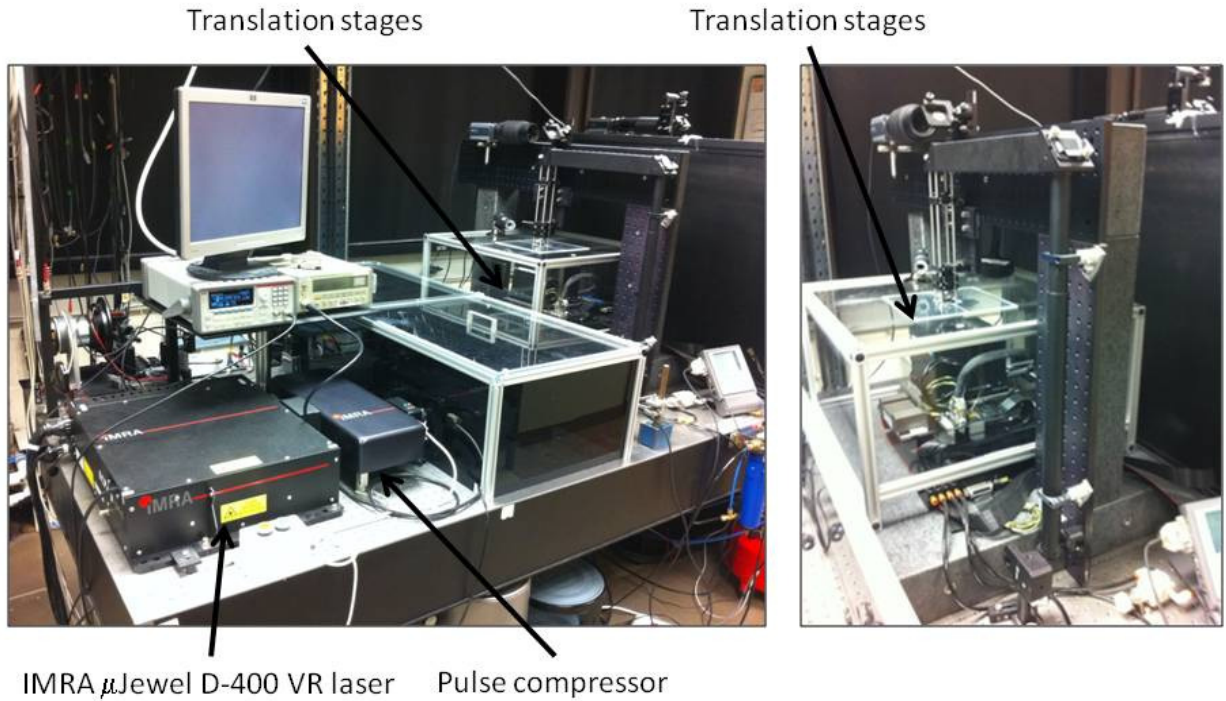

Figure 2. Photographs of the ULI manufacturing set-up. The picture on the left shows the entire ULI facility, the picture on the right shows a close-up of the translation stages and the granite arch which holds the optics used to manoeuvre the ULI laser onto the translation stages from above.

## 4. GRATING TESTS

The prototype ULI volume phase gratings are shown in Figure 3 with the fused silica gratings shown on the left and the GLS gratings on the right. The gratings are numbered as indicated in Figure 3. The fused silica substrate clearly shows the presence of 14 out of the 16 gratings whilst the GLS substrate shows 13 gratings. The missing gratings are ones where the laser power is not sufficient to produce a visible change of the refractive index and hence a grating. The yellow appearance of the GLS grating is due to the natural colour of the substrate material.

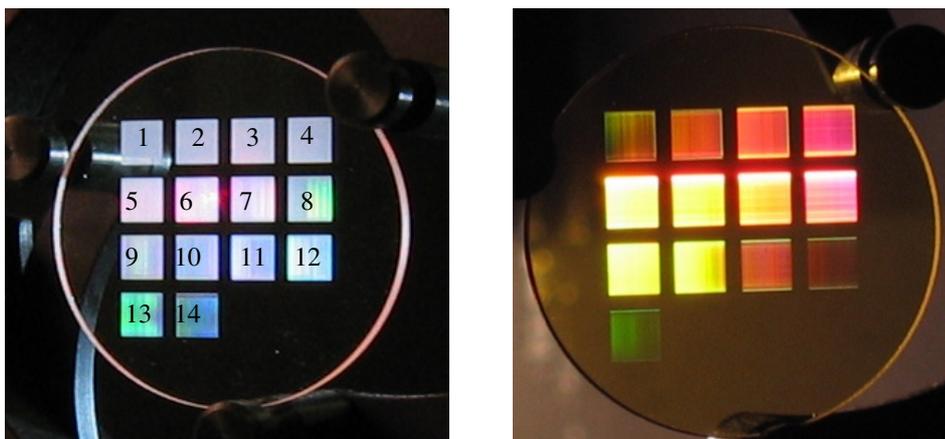

Figure 3. Pictures of the fused silica grating prototypes (left) and GLS grating prototypes (right).

For each of the gratings seen in Figure 3, the following performance parameters were measured: refractive index modulation, efficiency as a function of diffraction order and angle of incidence, and the integrated scattered light. These measurements were all performed at a wavelength of 633 nm with illumination provided by an intensity stabilised Helium-Neon laser.

### 4.1 Refractive index modulation

The refractive index modulation of the ULI grating structure was estimated from interferometric measurements of the grating and surrounding substrate. A 5 mm collimated beam from a commercial FISBA interferometer was used to measure the transmitted wavefront error of the diffraction grating and surrounding substrate. The interferometer was slightly tilted with respect to the grating to allow the observation of straight fringes. An obvious step occurs in the fringe pattern at the refractive index boundary between the substrate and the grating. From the fringe spacing and the size of the step it is straightforward to calculate the difference in optical path depth caused by the grating which in turn can be used to calculate the refractive index of the ULI modified region.

The results of the interferometric tests on the fused silica gratings showed an optical path depth of up to 50 nm and up to 100 nm for the GLS gratings. The depth of the grating structure is assumed to be 52 μm and 42 μm for the fused silica and GLS gratings respectively. The measured wavefront error indicates the average refractive index modulation. To calculate the index modulation within the inscribed grating line it is assumed that the total inscribed area occupies one third of the total grating area. The refractive index modulation of fused silica is therefore estimated to be in the range 0.002 – 0.003, similar to previously published values [17]. The refractive index modulation for GLS is somewhat higher at 0.007 – 0.012.

### 4.2 Grating efficiency

The diffraction efficiency of the prototype ULI gratings was measured using the simple test set-up shown Figure 4. The experimental set-up consists of a laser light source, a blocking filter, the grating under test, and a power meter. As the gratings are only 3 mm in size it was important to have a small collimated beam, so a Helium-Neon laser with a beam size of approximately 1 mm was selected. The grating itself is mounted on a stage with three degrees of motion to permit X-Y movement to select a grating, and rotation to the desired angle of incidence. The intensity of the laser output is recorded using a detector consisting of an integrating sphere, silicon photodiode, and power meter. The integrating sphere is mounted on an adjustable base to allow it to be moved to capture the light from the various diffraction orders.

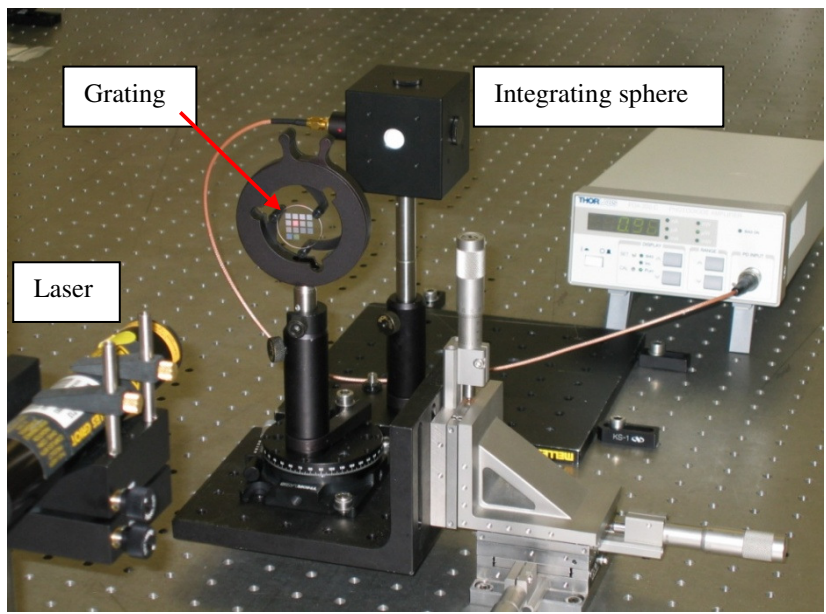

Figure 4. Pictures of the grating efficiency test experiment.

For each grating, on both the fused silica and GLS substrates, the following measurements were performed: transmittance of the bare substrate, efficiency of -2, -1, 0, +1, +2 orders at 0 degrees angle of incidence, and efficiency of -2, -1, 0, +1, +2 orders at 6.5 degrees angle of incidence (the blaze angle at 633 nm).

The best performing fused silica gratings were found to be 5 – 9, corresponding to laser pulse energies of ≈ 490 nJ → 340 nJ. For the GLS gratings numbers 8 – 10, corresponding to laser pulse energies of ≈ 100 nJ → 80 nJ, were found to have the highest amount of diffracted light. These gratings were tested in more detail to determine their diffraction efficiency as a function of angle of incidence. A plot of the first and zeroth order efficiency of GLS grating number 8, versus angle of incidence, is shown in Figure 5. The peak efficiency occurs at an angle of approximately 6.5 degrees consistent with the predicted blaze angle at 633 nm. The peak first order efficiency of 49 % is the absolute efficiency and includes absorption and reflection losses caused by the GLS substrate. The external transmittance of the GLS substrate is 69 %. If the grating efficiency is corrected for the transmittance of the GLS this implies the peak diffraction efficiency is 71 %.

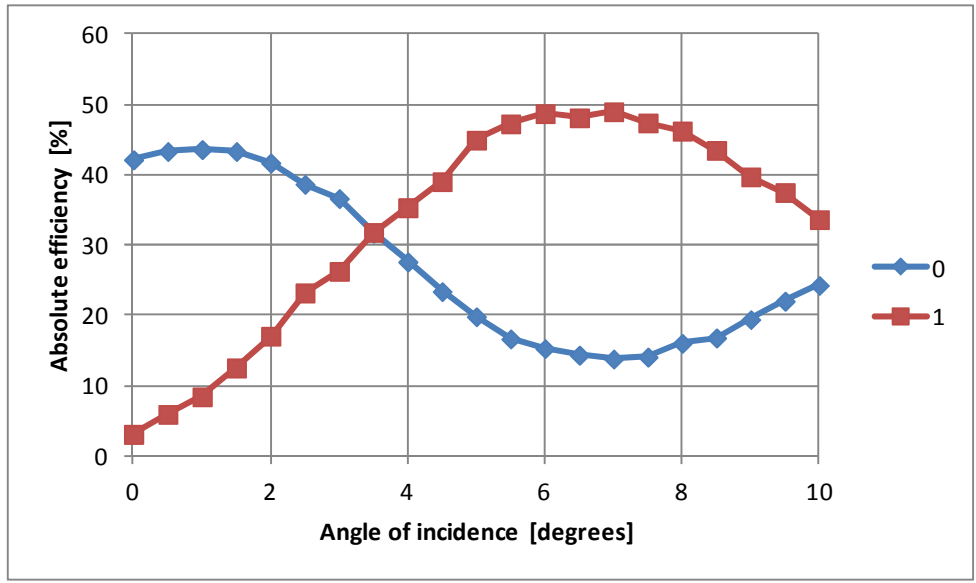

Figure 5. Plot of efficiency, for zeroth (blue line) and first (red line) diffraction orders, versus angle of incidence for GLS grating number 8 measured at 633 nm.

### 4.3 Scattered light

Both the fused silica and GLS gratings exhibit ghost images between the diffraction orders, as seen in Figure 6 for the fused silica grating. The origin of these ghost images is still under investigation.

The amount of scattered light was determined by measuring the total integrated transmittance of the grating and subtracting the known efficiency of the diffraction orders. The total integrated transmittance includes the contribution from all of the diffraction orders, ghosts, and the halo of scattered light. The total integrated transmittance is measured by placing the entrance aperture of the integrating sphere, shown in Figure 4, within a few millimetres of the grating such that most of the output beam is captured.

The scattered light performance of the two substrates was found to be quite different. The fused silica gratings exhibited large amounts of scattered light, approximately 40 %, whilst the GLS gratings produced less than 5 %. An image of the laser beam passing through the fused silica grating is shown in Figure 6. The diffraction orders and halo of scattered light have been indicated. Although only grating number 3 is illuminated by the laser beam, all of the gratings can be seen to be glowing, and light can also be seen coming from the edges of the substrate. The fused silica gratings scatter approximately 10 % of the incident light within the substrate. The remaining 30 % is scattered in the direction of the diffraction orders, and forms a halo of scattered light which can be seen on paper screen shown in Figure 6. These results for fused silica are similar to those described elsewhere [20]. It is expected that scattered light performance can be improved by a suitable adjustment to the ULI process.

The scattered light performance of the fused silica gratings implies that the ULI process is causing micro and nano-scale damage within the glass and these defects are causing scattered light to be distributed over large angles. This causes the fused silica ULI gratings to appear opaque (Figure 3 left). The ULI process in GLS performs much better, the gratings appear transparent (Figure 3 right), and the scattered light is primarily located in ghosts and the stripe of scattered light located between the grating orders. The GLS grating scattered light performance is not dissimilar to that of a conventional ruled surface relief grating.

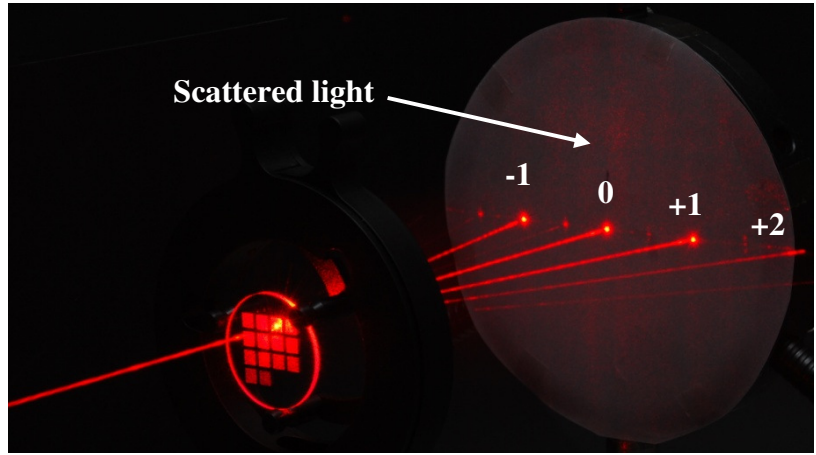

Figure 6. Image of a 633 nm laser beam illuminating the fused silica ULI grating prototype.

### 4.4 Cryogenic performance

If this grating technology is to be used in the thermal infrared it is important to demonstrate that the gratings can survive cooling to cryogenic temperatures and that their performance is unaffected by cooling. Both the fused silica and GLS gratings have been subjected to a number thermal cycles, from room temperature to 20 K, and show no visible degradation. The efficiency at cryogenic temperatures has not yet been characterised but is unlikely to be affected by cooling. One of us (RRT) has recently investigated the cryogenic operation of waveguide structures manufactured using ULI and found that the waveguide structures survived cooling to 4 K.

## 5. SUMMARY & CONCLUSIONS

A summary of the performance of the fused silica and Gallium Lanthanum Sulphide ULI gratings is given in Table 3. It is clear that the GLS substrate represents a better material for the production of volume phase gratings with the ULI process. The performance of the GLS gratings is such that their use in astronomical spectrometers should be considered, particularly if the performance demonstrated here in the visible can be obtained at infrared wavelengths.

Table 3. Summary of ULI volume phase grating test results.

| Test parameter | Fused Silica | Gallium Lanthanum Sulphide |
|---|---|---|
| Maximum change in refractive index | 0.002 – 0.003 | 0.007 – 0.012 |
| GSolver efficiency prediction | 65 % | 50 % |
| Maxiumum absolute efficiency First order at 633 nm | 37 % | 49 % |
| Substrate transmittance at 633 nm | 93 % | 69 % |
| Maximum relative efficiency First order at 633 nm | 40 % | 71 % |
| Total integrated diffracted light | 50 – 60 % | 95 % |
| Total scattered light at 633 nm | ~ 40 % | < 5 % |
| Ghost images | Between orders | Between orders |
| Gratings survive cooling to 20 K | Yes | Yes |

The initial results from the ULI volume phase gratings have been extremely encouraging. The next steps in the development of this technology are to demonstrate the operation of a GLS grating in the infrared and to manufacture a larger grating, with the aim of testing it in an astronomical instrument. There is clearly more work to do to investigate the optimum ULI processing parameters to achieve greater refractive index modulation.

The manufacture of a large astronomical grating represents quite a challenge for the ULI equipment, particularly the stability requirements on the grating positioning stage. A modestly sized 100 mm by 100 mm grating, with 300 lines per mm, represents over 3 km of inscription per layer, and perhaps 50 layers might be needed for the grating to operate efficiently in the infrared. At a scan speed of 10 mm per second this example grating would take approximately half a year to manufacture. Alternative manufacturing strategies, e.g. [6], are therefore needed to be able to produce large astronomically useful ULI gratings within reasonable, and economic, timescales.

The robust nature of the ULI grating structures, the ability to manufacture them in a wide range of materials, and their high performance, suggests they will also suit a variety of applications outside astronomy such as, hyper-spectral imaging satellite sensors and small laboratory spectrometer systems.

## ACKNOWLEDGMENTS

This work was funded by the STFC through RRT's Fellowship (ST/H005595/1) and the EPSRC (EP/G030227/1). RRT thanks AKK for access to the ULI facility. DL acknowledges the work of senior honours students Ellen Milne, Angus Irvine, and Creag Carson and the financial support of STFC's Centre for Instrumentation.